\documentclass[prl,twocolumn,showpacs,superscriptaddress,nofootinbib,floatfix,10pt]{revtex4-1} %notitlepage,
\usepackage[utf8]{inputenc}
\usepackage{color}
\usepackage{amsfonts,amsmath,amssymb}
\usepackage{longtable,booktabs}
\usepackage{graphicx}
\usepackage{bm}
\usepackage{hyperref}
\usepackage{epstopdf}
\usepackage{mathtools}
\usepackage{gensymb}

\usepackage[utf8]{inputenc}
\usepackage{graphicx}
\usepackage{multirow}
\usepackage{comment}
\usepackage{pifont}
\usepackage{slashed}
\usepackage{bm}
\usepackage{tabularx}
\usepackage{mhchem}
\usepackage{pifont}

\def\c{\bar c}

\def\jp{J\!/\psi}
\def\M{\overline M}

\def\*{^{(*)}}

\def\x4c/{$X_{cc\bar c \bar c }$}
\def\>{\big>}
\def\<{\big<}
\def\|{\big\vert}

\def\be{\begin{equation}}
\def\ee{\end{equation}}
\def\ba{\begin{eqnarray}}
\def\ea{\end{eqnarray}}

\begin{document}

\title{Predictions for the scalar partner of the LHC tetraquark $X(6600)$}

\author{Muhammad Naeem Anwar}\email[E-mail address: ]{m.n.anwar@swansea.ac.uk}
\author{Timothy J. Burns}\email[E-mail address: ]{t.burns@swansea.ac.uk}
\affiliation{Centre for Quantum Fields and Gravity, Department of Physics, Swansea University, Singleton Park, Swansea, SA2 8PP, United Kingdom}

\begin{abstract}
We consider how the recent CMS measurements of the masses and quantum numbers of $X(6600)$, $X(6900)$ and $X(7100)$ can help to reveal the internal structure of these apparent $cc\bar{c}\bar{c}$ tetraquark states. The measured $J^{PC} = 2^{++}$ quantum numbers of $X(6600)$ are consistent with our previous prediction, and imply the existence of lighter $0^{++}$ partner $X(6400)$ which also decays to $\jp\jp$. There may already be indications for this scalar partner in the recent CMS data fits, which include a Breit-Wigner peak with mass around 6400~MeV. We give predictions for the masses and decay properties of the scalar and other partner states, which are key experimental tests to discriminate between quark and diquark models. We urge closer experimental scrutiny in this mass region, to establish an S-wave multiplet  of $cc\c\c$ states (the first of its kind), leading to a breakthrough in exotic hadron research and our understanding of exclusively heavy quark exotics.

\end{abstract}

\date{\today}

\maketitle

{\it \textbf{Introduction}}.
The experimental observations of all-charm tetraquarks have opened new directions for exotic hadron spectroscopy, partly because the dynamics of exclusively heavy quark systems are relatively ``cleaner" in comparison to systems with light quarks. The experimental status, summarised in ref.\,\cite{CMStalk2026}, is that $cc \c \c $ states have been observed at three different LHC experiments, initially at LHCb~\cite{LHCb:2020bwg}, then ATLAS~\cite{ATLAS:2023bft}, and more recently CMS~\cite{CMS:2023owd,CMS:2026tiu}. 

Particularly significant recent developments have come from the CMS Collaboration. Amplitude analysis of the three resonance structures $X(6600)$, $X(6900)$ and $X(7100)$ in the $\jp \jp$ channel, based on Run 2 data, favours all three states having the same $J^{PC}$ quantum numbers, with $2^{++}$  preferred over other possibilities \cite{CMS:2025fpt}. Additionally, with an updated analysis of combined Run 2 and Run 3 data, CMS have given refined measurements of their masses and decay widths and argued, on the basis of Regge trajectories, that the higher mass resonances are likely to be the radial excitations of the lower one \cite{CMS:2026tiu}. Meanwhile, the ATLAS Collaboration has observed $X(6900)$ decaying into $\jp \psi(2S)$ and extracted the ratio of the partial decay widths between the $\jp \psi(2S)$ and $\jp \jp$ channels~\cite{ATLAS:2025nsd}. 

An intriguing feature, which may be crucial in developing a theoretical understanding of these states, is that all three experiments show indications of another possible structure around 6400~MeV \cite{LHCb:2020bwg,ATLAS:2023bft,CMS:2023owd,Giron:2020wpx}, and notably this has become even more prominent in the recent CMS analysis with more data \cite{CMS:2026tiu}. The experimental properties of this putative state are, however, not yet established. At ATLAS \cite{ATLAS:2023bft}, one of two possible fits to experimental data includes a Breit-Wigner to describe this enhancement, with mass $m=6410\pm 80$~MeV. The recent CMS analysis \cite{CMS:2026tiu} also includes a Breit-Wigner to describe the enhancement around 6400~MeV, although curiously it is attributed to experimental background. We will argue below, on the basis of the experiment data and phenomenological considerations, that the enhancement around 6400~MeV is most likely a (genuine) scalar state and warrants closer experimental scrutiny.

Prior to these experimental developments, we explored the phenomenology of $cc\c\c$ states in refs. \cite{Anwar:2023fbp,Anwar2023}. In the current paper we will update and extend on this previous work, taking account of the recent experimental developments at CMS. One of our key predictions was that $X(6600)$ and its apparent partner state $X(6400)$ belong to an S-wave multiplet of $cc\c\c$ tetraquarks, with $2^{++}$ and $0^{++}$ quantum numbers, respectively. Our prediction of $2^{++}$ quantum numbers for $X(6600)$ has now been confirmed at CMS, and we therefore urge further experimental study of the 6400~MeV enhancement, to establish whether it corresponds to the $X(6400)$ state whose properties we predicted in our previous work. Using the refined measurements of the $X(6600)$ mass \cite{CMS:2026tiu}, and mass relations we derived in previous work~\cite{Anwar:2023svj}, below we give our predictions for the masses of partner states, including (but not only) the scalar $X(6400)$. In addition we review and update our predictions for the relative decay branching fractions into different final states, and highlight how these can discriminate between quark and diquark models.

%%%%%%%%%%%%%%%%%%%%%%%%%%%%%%%%%%%%%%%~~~NEW SECTION~~~%%%%%%%%%%%%%%%%%%%%%%%%%%%%%%%%%%%%%%%%%%
%\vspace{0.2em}
{\it \textbf{Mass splittings and multiplets}}.
Quark and diquark models predict a rich spectrum of $cc\c\c$ states, organised into multiplets whose states have the same radial and orbital quantum numbers $n$ and $L$, but different spin $S$ or total angular momentum $J$. The composition of these multiplets varies depending on the assumed degrees of freedom: constituent quark models have a richer spectrum than diquark models, due to additional colour configurations and a larger number of internal coordinates subject to radial or orbital excitation. 

Models vary considerably in their predicted masses, as summarised for example in ref.~\cite{Anwar:2023fbp}. But a feature common to all models is that the splittings within a multiplet are considerably smaller than the range of masses spanned by the established experimental states $X(6600)$, $X(6900)$ and $X(7100)$. Consequently there is no plausible scenario in which all these experimental states belong to a single multiplet.

In their analysis of the $\jp\jp$ spectrum in the region of $X(6600)$, $X(6900)$ and $X(7100)$, CMS find that incoherent Breit-Wigner distributions cannot adequately fit the data, whereas allowing for interference yields a much-improved fit. On this basis they conclude that the three states have the same $J^{PC}$ quantum numbers~\cite{CMS:2023owd,CMS:2026tiu}, and with this assumption, their analysis of angular distributions strongly prefers $2^{++}$ over other possibilities \cite{CMS:2025fpt}. The states could have different $n$ and/or $L$ quantum numbers, and the particular scenario advocated by CMS is that all three are $n\ce{^5S_2}$ states, with radial quantum numbers $n=1,2,3$ or $n=2,3,4$. 

Another possibility, which deserves a little more attention and careful consideration in the experimental analysis, is that the states could have different $J$ quantum number. A particularly interesting possibility is that one or more of the states (including the possible structure around 6400~MeV) could be $0^{++}$; in all models, radial excitations $n\ce{^5S_2}$ (as advocated by CMS) are accompanied by $n\ce{^1S_0}$ partners. These $0^{++}$ states can also decay into $\jp \jp$, albeit with a relatively smaller decay rate in comparison to a $2^{++}$ state~\cite{Anwar:2023fbp,Becchi:2020uvq,Wang:2023kir} (we discuss this further below). Hence, relaxing the assumption of interfering states and allowing different $J$ values in the fits would be insightful for spectroscopy and understanding their quark structure.

%%%%%%%%%%%%%%%%%%%%%%%%%%%%%%%%%%%%%%%~~~NEW SECTION~~~%%%%%%%%%%%%%%%%%%%%%%%%%%%%%%%%%%%%%%%%%%
%\vspace{0.5em}
{\it \textbf{Tensor state $\bm{X(6600)}$}}.
In our previous work~\cite{Anwar:2023fbp} we concentrated on $X(6600)$, and its possible partner state ``$X(6400)$'', for which there was some evidence at ATLAS. On the basis of their masses and relative prominence in $\jp\jp$ decays, we argued that these are  $1\ce{^5S2}$ and $1\ce{^1S_0}$ states, respectively, leading to predictions $2^{++}$ and $0^{++}$ for their quantum numbers. Our $2^{++}$ prediction for $X(6600)$ has now been confirmed by CMS~\cite{CMS:2025fpt}, in the analysis described above. As for $X(6400)$, our $0^{++}$ prediction remains to be tested experimentally, since this state was not studied explicitly in the CMS analysis; we comment further on this point below.

The rationale for assigning $X(6400)$ and $X(6600)$ to a $1\ce S$ multiplet is partly because their masses are consistent with a naive estimate~\cite{Anwar:2023fbp} derived from the $\Xi_{cc}^{++}$ mass $3621.40 \pm 0.78 $~MeV~\cite{LHCb:2017iph}, and broadly in agreement with the predictions of a range of quark and diquark models; see for example refs~\cite{liu:2020eha,Lloyd:2003yc,Ader:1981db,Lin:2024olg,An:2022qpt,Deng:2020iqw,Galkin:2023wox}, and our summary of predictions in~\cite{Anwar:2023fbp}. 

Within the 1S multiplet, the specific assignment of $0^{++}$ and $2^{++}$ quantum numbers to $X(6400)$ and $X(6600)$, respectively, is due to the mass ordering and splitting \cite{Anwar:2023svj,Anwar:2023fbp}. In diquark models there are three states in the $1\ce S$ multiplet, with quantum numbers  $0^{++}$, $1^{+-}$ and $2^{++}$, in increasing order of mass. Only those with $C=+1$ can decay to $\jp\jp$ so, on the basis of the mass ordering, we assigned $0^{++}$ and $2^{++}$ quantum numbers to $X(6400)$ and $X(6600)$. In quark models an additional $0^{++^\prime}$ state is predicted, higher in mass than the others. In this case there are three states which could decay into $\jp\jp$, so the assignment of $0^{++}$ and $2^{++}$ quantum numbers to $X(6400)$ and $X(6600)$ is not automatic, but is based on the mass splittings: quark models predict $\sim 150 \pm 15$ MeV separation between the ground state $0^{++}$ and the $2^{++}$~\cite{Anwar:2023fbp}, comparable to the spacing between the CMS peaks \cite{CMS:2026tiu}; the splitting between $2^{++}$ and the higher-lying $0^{++\prime}$ is predicted to be considerably smaller.

As well as spectroscopy, our quantum number assignment is also supported by decays. We previously found that in both quark and diquark models the decay $2^{++}\to\jp\jp$ is enhanced in comparison to $0^{++}\to\jp\jp$~\cite{Anwar:2023fbp,Anwar2023}, a result also found in other approaches~\cite{Becchi:2020uvq,Wang:2023kir,Maiani:2020pur,Agaev:2026mif,Zhang:2020hoh, Celiberto:2025ziy}. This is consistent with the $2^{++}$ assignment for $X(6600)$, noting the prominence of the corresponding peak in $\jp\jp$ data. Related to this, a recent analysis of $\jp\jp$ scattering in lattice QCD also prefers $2^{++}$ quantum numbers for $X(6600)$~\cite{Li:2025ftn,Li:2025vbd}.

As mentioned above, CMS has proposed that $X(6600)$, $X(6900)$ and $X(7100)$ are $n\ce{^5S_2}$ states, with radial quantum numbers $n=1,2,3$ or $n=2,3,4$, respectively \cite{CMS:2026tiu}. Our $1\ce{^5S_2}$ assignment for $X(6600)$ agrees with the first of these scenarios. We did not consider the higher states $X(6900)$ and $X(7100)$ in our previous work, because of the difficulty in establishing model-independent results in multiplets of radial excitations (Note that the ATLAS did not observe $X(7100)$ in the $\jp \psi(2S)$ channel~\cite{ATLAS:2025nsd} where CMS has claimed its evidence~\cite{CMS:2025vnq,CMS:2026tiu}.) It is worth emphasising that classifying radial excitations with a single quantum number, as proposed by CMS, is only legitimate where there is a single radial coordinate, namely in the diquark model (see for example refs.~\cite{Ali:2019roi,Galkin:2023wox,Lin:2024olg}). The situation is more complicated in quark models, for which tetraquarks have three internal coordinates, meaning the spectrum of excited states is richer (and denser) than that of the ground state multiplet, presenting a challenge both for theory and experiment; see refs.~\cite{Wang:2019rdo,Lu:2020cns,liu:2020eha} for examples of quark model spectra with excited states. For this reason in ref.~\cite{Anwar:2023fbp} we did not generalise our results for the ground state multiplet (in the quark model) to higher-lying multiplets.

%%%%%%%%%%%%%%%%%%%%%%%%%%%%%%%%%%%%%%%~~~NEW SECTION~~~%%%%%%%%%%%%%%%%%%%%%%%%%%%%%%%%%%%%%%%%%%
%\vspace{0.5em}
{\it \textbf{Predicted scalar state $\bm{X(6400)}$}}.
Although the CMS analyses \cite{CMS:2023owd,CMS:2025fpt,CMS:2026tiu} do not extract the properties of the $X(6400)$ structure observed at ATLAS \cite{ATLAS:2023bft}, there is apparent structure around this mass in the CMS data, which has become more prominent in the new combined Run 2+3 data~\cite{CMS:2026tiu} (the only analysis of all-charm resonances with combined data). The enhancement is sufficiently prominent that in order to describe data in this mass region, the CMS fits include a Breit-Wigner amplitude (referred to as BW0) characterised by a mass around 6400 MeV. Curiously, however, despite having similar prominence and a comparable width to the amplitudes describing $X(6600)$, $X(6900)$ and $X(7100)$,  this feature is attributed to experimental background, rather than a genuine state. It is striking to notice that the resulting background is unusually sharply peaked.

For the sake of argument, let us assume that the Breit-Wigner around 6400 MeV describes a $cc\c\c$ state, $X(6400)$. Although we have no direct experimental information on its quantum numbers, we can still infer something useful from the CMS analysis. As mentioned above, a good description of the invariant mass distributions requires interference between the amplitudes for $X(6600)$, $X(6900)$ and $X(7100)$, indicating these three states have the same quantum numbers, with $2^{++}$ strongly preferred. In these fits, the Breit-Wigner describing $X(6400)$ is added incoherently, indicating that the data do not require interference with the other states. On this basis we can infer that the quantum numbers of $X(6400)$ are likely to be distinct from those of the other states, which is consistent with our prediction of $0^{++}$ for this state.

In any case, a generic feature of quark and diquark models is that a $2^{++}$ state should be accompanied by a $0^{++}$ partner, with lower mass. Since CMS has confirmed the $2^{++}$ quantum numbers for $X(6600)$, it is very natural to expect a $0^{++}$ partner nearby, and in this context the apparent structure around 6400 MeV is tantalising and warrants closer scrutiny. This peaking structure emerges exactly at the location where the $0^{++}$ state is expected from the quark models and QCD sum rules~\cite{liu:2020eha,Lloyd:2003yc,Ader:1981db,Anwar:2017toa,An:2022qpt,Deng:2020iqw,Zhang:2020xtb,Yang:2020wkh,Wu:2022qwd}, approximately $\sim 150 \pm  15$ MeV below the tensor state \cite{Anwar:2023fbp}. As mentioned above, the decay $0^{++}\to\jp\jp$ is expected to be suppressed relative to $2^{++}\to\jp\jp$, which is qualitatively consistent with the CMS data, where the peak associated with $X(6400)$ is less prominent compared to $X(6600)$.

As in our previous work~\cite{Anwar:2023fbp,Anwar2023}, we suggest further experimental study of the enhancement around $X(6400)$. Measurements of its mass and width can be confronted with those obtained at ATLAS~\cite{ATLAS:2023bft}, and compared to theoretical predictions including those in our previous work~\cite{Anwar:2023fbp}, updated below. It would be particularly insightful to include the Breit-Wigner associated with the structure around 6400~MeV (which CMS refers to as BW0) in an angular analysis together with $X(6600)$, $X(6900)$ and $X(7100)$, and relaxing the assumption of the same quantum numbers for all these states, to determine their quantum numbers. In particular it would be useful to include the lowest peak (referred to as BW0 in the CMS analyses) incoherently with the higher three coherent BWs and extract the parameters of BW0 to confront with ATLAS and theoretical predictions. Experimental observation of several states in the S-wave multiplet would be very insightful to constrain theoretical models, in particular to discriminate between competing theoretical models, as discussed later.

There may also be hints in the data of multiple, narrower peaks around $6400$ MeV, an intriguing possibility which could be resolved in future analyses with higher statistics. This is reminiscent of other examples in exotic hadron spectroscopy, where high resolution study of nearby peaks reveals two resonances (possibly with different quantum numbers), for example the hidden-charm pentaquarks~\cite{LHCb:2019kea,Burns:2019iih,Du:2019pij,Du:2021fmf,Burns:2022uiv,Nakamura:2021qvy,Kuang:2020bnk}, and the so-called $Y$ states with $J^{PC} = 1^{--}$~\cite{Wang:2025clb,Bai:2026atm,Anwar:2021dmg}.

%%%%%%%%%%%%%%%%%%%%%%%%%%%%%%%%%%%%%%%~~~NEW SECTION~~~%%%%%%%%%%%%%%%%%%%%%%%%%%%%%%%%%%%%%%%%%%
%\vspace{0.5em}
{\it \textbf{Partner states}}.
As well as the $0^{++}$ partner discussed in the previous section, the $2^{++}$ state $X(6600)$ should be accompanied by an axial state $1^{+-}$, and possibly also a heavier scalar state $0^{++\prime}$, depending on the model ansatz. In this section we give predictions for the masses of these partners.

In ref. \cite{Anwar:2023svj} we derived mass formulae, in quark and diquark models, for S-wave tetraquarks having either one or two quark flavours. Specialising to $cc\c\c$ tetraquarks~\cite{Anwar:2023fbp}, the formulae are expressed in terms of the multiplet centre-of-mass $\M$ (which is either fit to data or derived from a model) and the quark-quark and quark-antiquark couplings $C_{cc}$ and $C_{c\c}$ (fit to the spectra of baryons and mesons). The formulae can also be conveniently expressed in terms of the ratio of couplings $R=\dfrac{C_{c\c}}{C_{cc}}$, with the symmetry limit $R=1$ often assumed in model calculations.

In both quark and diquark models, the masses of the axial $(1^{+-})$ and tensor ($2^{++}$) are
\begin{align}
    M_1&=\M+\frac{16}{3}C_{cc}(1-R),\label{eq:m1}\\
    M_2&=\M+\frac{16}{3}C_{cc}(1+R). \label{eq:m2}
\end{align}
In diquark models, there is a single scalar $(0^{++})$ with mass
\begin{align}
M_0&=\M+\frac{16}{3}C_{cc}(1-2R), \label{eq:mf0} 
\end{align}
whereas in quark models there are two scalars ($0^{++}$ and $0^{++\prime})$ with masses
\begin{align}
    M_0&=\M+\frac{4}{3}C_{cc}\left(5-4R-\Delta\right),\label{eq:m0}\\
    M_0'&=\M+\frac{4}{3}C_{cc}\left(5-4R+\Delta\right),\label{eq:m0p}
\end{align}
where $\Delta=\sqrt{232R^2+8R+1}$.

In order to update our previous predictions for tetraquark masses, we fix $M_2$ to the $X(6600)$ mass measured at CMS~\cite{CMS:2026tiu}, and compute the masses of axial vector and scalar states in the quark and diquark models. Working first in the symmetry limit $R=1$ (i.e., $C_{cc}=C_{c\c}$), we adopt $ C_{cc} = 5.0 \pm 0.5$~MeV, on the basis of various quark model extractions~\cite{Deng:2020iqw,Buccella:2006fn,Liu:2019zoy}. The masses obtained in this way are summarised in Table~\ref{table:masses}. 

\begin{table}[h!]
\centering
\begin{tabular}{l|c|c}
\hline\hline 
$J^{PC}$\qquad\qquad & Quark Model (MeV) & Di-quark Model (MeV) \\ [1ex] 
\hline
$0^{++}$ & 6443 $\pm$ 19 &  6513 $\pm$ 16 \\
$1^{+-}$& 6540 $\pm$ 16  & 6540 $\pm$ 16 \\ 
$2^{++}$(Input \cite{CMS:2026tiu})& 6593 $\pm$ 15 & 6593 $\pm$ 15 \\
 $0^{++'}$ & 6650 $\pm$ 19 & \ding{55} \\ [1ex]
\hline
\end{tabular}
\caption{Predicted masses of $S$-wave $cc \bar c \bar c$ states in the quark and diquark models using our previous work~\cite{Anwar:2023fbp,Anwar:2023svj}, having fixed the tensor ($2^{++}$) mass to the CMS value~\cite{CMS:2026tiu}, with $C_{cc}=5.0\pm 0.5$~MeV and $R=1$. The \ding{55} indicates that the state does not exist in the diquark model, as explained in the text.}
\label{table:masses}
\end{table}

The predicted spectra can be confronted with future experimental data as a test of theoretical models. In particular for the light scalar ($0^{++}$) there is a signficant difference (around $70$~MeV) between the predictions of quark and diquark models. Measuring the scalar mass would therefore help to resolve the internal structure of $cc \bar c \bar c$ systems, discriminating between quarks and diquarks as the relevant degrees of freedom. Similarly the discovery of an additional, heavier scalar ($0^{++\prime}$) would indicate quark (rather than diquark) degrees of freedom. The mass of the axial ($1^{+-}$) is the same in both models, and is a direct test of model parameters.

\begin{figure*}[t!]
   \centering
    \includegraphics[width=0.8\textwidth]{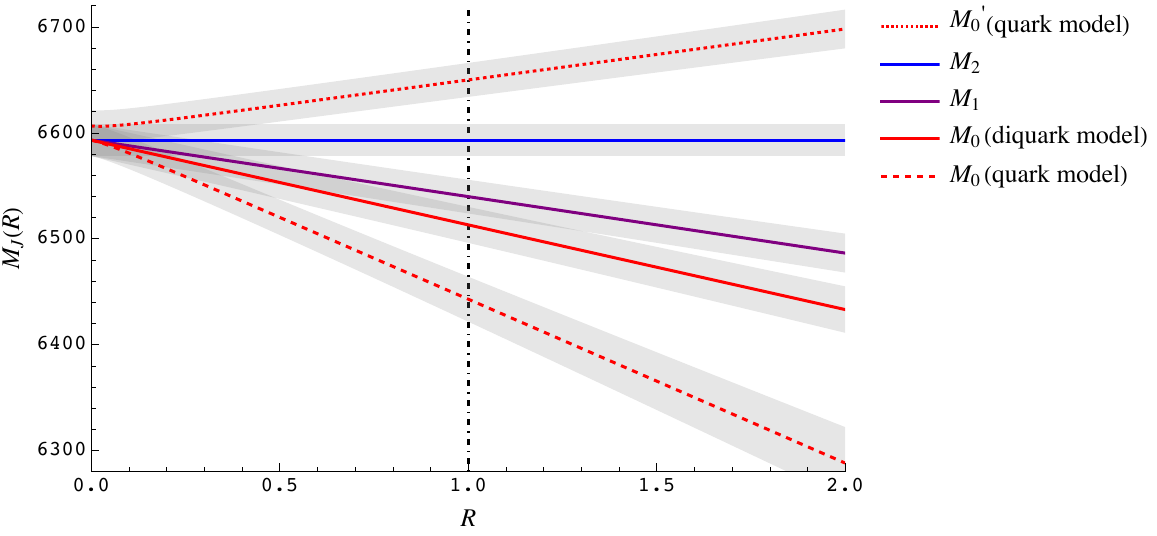}
   \caption{Mass predictions of S-wave $c\c\c\c$ states in the quark and diquark model as a function of $R$, with $C_{cc}=5.0\pm 0.5$~MeV.}
    \label{massplot}
\end{figure*}

To establish the $R$-dependence of our predictions, we fix $C_{cc}=5.0 \pm 0.5$~MeV but allow $C_{c\c}$ (hence $R$) to vary. The results are shown in Fig~\ref{massplot}. Having the $2^{++}$ mass fixed, the partner masses have mild $R$ dependence, and notably the predictions for the scalar mass $M_0$ in quark and diquark models are well-separated away from $R\to 0$ limit. 
%-------------------
%-------------------
Fig~\ref{massplot} also reveals some other features which result from the mass formulae above \cite{Anwar:2023fbp,Anwar:2023svj}. The ordering of masses is necessarily $M_0<M_1<M_2$ in diquark models, and $M_0<M_1<M_2<M_0'$ in quark models. (Note however that quark models with different assumptions to those of our previous work can yield different mass orderings, as mentioned in ref.~\cite{Anwar:2023fbp}.) 

Another interesting feature is that in the $R\to 0$ limit, the masses $M_0$, $M_1$ and $M_2$ become degenerate (revealed in Fig~\ref{massplot}), which is a known feature of the diquark model~\cite{Giron:2020wpx,Ali:2019roi}. Notice also that the quark and diquark model predictions for the scalar $M_0$, which are distinct for $R\ne 0$, converge as $R\to 0$; this is physically reasonable, since the concept of effective diquark degrees of freedom makes sense if $c\c$ interactions are negligible in comparison to $cc$ and $\c\c$ interactions \cite{Anwar:2023svj}. However the heavier scalar $M_0'$ in the quark model remains somewhat higher than the other three masses even as $R\to 0$.

It is interesting to consider how the mass predictions in different models relate to our proposal, discussed above, that the enhancement in the $\jp\jp$ data around 6400~MeV corresponds to the scalar $0^{++}$. Although both ATLAS and CMS include a Breit-Wigner in their fits to describe this structure, only ATLAS quotes the corresponding mass, $6410\pm 80$~MeV \cite{ATLAS:2023bft}. With reference to Fig~\ref{massplot}, for ``natural'' values of $R$ near the symmetry limit ($R=1$), the ATLAS measurement is consistent with the quark model, but not the diquark model. While clearly this supports the quark model description, a more precise experimental measurement would strengthen this conclusion.

Of course this analysis assumes that the couplings in tetraquark systems can be fixed by comparison to the corresponding couplings in mesons and baryons, which may not be valid. An alternative approach is to treat the couplings $C_{cc}$ and $C_{c\c}$ (or $C_{cc}$ and $R=C_{c\c}/C_{cc}$) as free parameters, which can be fit to reproduce experimental tetraquark mass splittings. A convenient implementation of this approach is to examine linear relations among the masses of tetraquarks, derived from linear combinations of the mass formulae above in which $C_{cc}$ (and $\overline M$) drop out~\cite{Anwar:2023svj, Anwar:2023fbp}. If the masses of any two states in the multiplet are measured in experiment, using these relations we can predict the masses of the other state (in diquark models), or the other two states (in quark models). With $M_2$ already known from CMS, how this plays out depends on which of the remaining states is identified next in experiment. 

We argued above that there may already be indications in experiment for the light scalar $0^{++}$. Once its mass $M_0$ is measured, then together with $M_2$ we can predict the mass of the axial in the diquark model,
\begin{align}
M_1=\frac{1}{3}\left(2M_0+M_2\right),\label{eq:diq:relation}
\end{align}
and the masses of both the axial and the heavier scalar in the quark model,
\begin{align}
M_1&=\frac{1}{8R+\Delta-1}\Big(8RM_0+\left(\Delta-1\right)M_2\Big)\,,\label{eq:relation1}\\
M_0'&=\frac{1}{8R+\Delta-1}\Big( (8R-\Delta-1)M_0+2\Delta M_2 \Big).
\end{align}
The diquark model prediction is independent of model parameters, while the quark model prediction depends only very weakly on $R$ in a reasonable range~\cite{Anwar:2023fbp}. The relations give a useful experimental test of models, since in this approach the predicted mass of the axial is very different for quark and diquark models; see for example Fig.~2 of ref.~\cite{Anwar:2023fbp}. (Note the distinction with the approach outlined above, where we found that with the same couplings, the models yield the same prediction for $M_1$; here instead we are fixing the masses $M_0$ and $M_2$, which effectively implies different couplings in the two models, hence distinct predictions for $M_1$.)

Another possibility is that, following $M_2$, the next mass to be discovered is that of the axial $M_1$. (We discuss the experimental prospects in the next section.) In this case the relations can be used to predict the masses of the scalar states. In diquark models there is one scalar, with mass
\begin{align}
M_0 & =\frac{1}{2}\left(3M_1 - M_2\right) \, , 
\label{eq:diq:relation}
\end{align}
whereas in quark models there are two scalars, with masses
\begin{align}
M_0 & =\frac{1}{8R}\Big((8R+\Delta-1) \, M_1 - (\Delta - 1) \, M_2 \Big) \, , \\
M_0' & =\frac{1}{8R}\Big( (8R-\Delta-1) \, M_1 + (\Delta + 1) \, M_2 \Big) \, . 
\label{eq:q:relation}
\end{align}
These mass relations have almost no model dependence, as tested for the several different variants of quark and diquark models~\cite{Anwar:2023svj}. The scalar state is well separated in both models.

Finally, note that for the diquark model, the mass relations above apply not only to the ground state 1S multiplet, but also the higher multiplets (radial excitations) such as 2S and 3S; hence they may be applied in future as tests of the CMS assignment that $X(6600)$, $X(6900)$ and $X(7100)$ are $n\ce{^5S_2}$ states, with $n=1,2,3$ or $n=2,3,4$. The situation is not so simple in quark models for which, as mentioned above, there is a richer spectrum of states in radially excited multiplets due to additional internal coordinates.

%%%%%%%%%%%%%%%%%%%%%%%%%%%%%%%%%%%%%%%~~~NEW SECTION~~~%%%%%%%%%%%%%%%%%%%%%%%%%%%%%%%%%%%%%%%%%%
%\vspace{0.5em}
{\it \textbf{Decays}}.
As well as the masses, the decay patterns also provide a direct and crucial test of the sub-structure of these states. In our previous paper \cite{Anwar:2023fbp}, we derived relations among partial decay widths for the S-wave decays of $cc\c\c$ tetraquarks into charmonium pairs, specifically $\jp\jp \, \{2^{++}, 0^{++(\prime)}\}$, $\eta_c\eta_c \, \{0^{++(\prime)}\}$ and $\jp\eta_c \, \{1^{+-}\}$. These assume the same spatial wavefunctions for the tetraquarks  (and, where applicable, the same spatial wavefunctions for $\jp$ and $\eta_c$), such that the relative amplitudes are determined by the overlap of spin and colour wavefunctions of the initial and final states. The dependence on hadron masses (via phase space factors) is insignificant, so in this paper we quote predictions using the masses assigned in our previous paper; these predictions can be updated once the masses of additional partner states are known. For the quark model the results depend on the mixing angle, for which we adopt the value ($\theta=35.6\degree$) derived assuming $R=1$; sensitivity to mixing angle is discussed in ref.~\cite{Anwar:2023fbp}.

As discussed, the quantum number assignment $n\ce{^5S_2}$ to the states at CMS implies spin-singlet partner states $n\ce{^1S_0}$ somewhat lower in mass, and these are likewise expected to decay into $\jp \jp$. However we find the decay $0^{++}\to\jp\jp$ is suppressed relative to the observed $2^{++}\to\jp\jp$ mode, which may explain the apparent suppression of the scalar partner states in the $\jp\jp$ data. For the ground state multiplet we predict the ratio $\Gamma(0^{++}\to\jp\jp)/\Gamma(2^{++}\to\jp\jp)$ is $0.19$ in the diquark model ($1/5$ before phase space factors), versus $0.073$ in the quark model. The suppression of the scalar decay in comparison to the tensor is also found in QCD sum rules~\cite{Wang:2023kir,Agaev:2026mif} and  NRQCD~\cite{Zhang:2020hoh,Feng:2023agq,Celiberto:2025ziy}, as well as other quark model calculations which, additionally, identify some other intriguing features distinguishing the scalar and tensor modes~\cite{Chen:2024orv}.

Models agree that although the light scalar is suppressed in $\jp\jp$, it decays prominently in $\eta_c\eta_c$~\cite{Anwar:2023fbp,Becchi:2020uvq,Wang:2023kir}. We find the ratio of partial widths $\Gamma(0^{++}\to\eta_c\eta_c)/\Gamma(0^{++}\to\jp\jp)$ is $4.3$ in the diquark model, versus $19$ in the quark model \cite{Anwar:2023fbp}. The $\eta_c\eta_c$ channel is a particularly ``clean" channel for the discovery of $0^{++}$ since for other states this mode is suppressed: for $2^{++}$ the decay is D-wave, and for  the heavier scalar $0^{++\prime}$ (in the quark model) the decay is suppressed by spin and colour overlaps. These results offer another solid test that the enhancement around $6400$ is the light scalar $0^{++}$; the prediction that it decays more prominently into $\eta_c\eta_c$ than $\jp\jp$ applies to both diquark models and quark models, although the enhancement of $\eta_c\eta_c$ is significantly stronger in the latter case. 

Finally we remark that the axial-vector $1^{+-}$ is expected to leave prominent signatures in $\eta_c\jp$ final state. Comparing to the observed decay for $X(6600)$ we find $\Gamma(1^{+-}\to\eta_c\jp)/\Gamma(2^{++}\to\jp\jp)=1.08$ in both quark and diquark models. An initial search by Belle found evidence for $e^+ e^- \to \eta_c\jp$ near the threshold~\cite{Belle:2023gln}, and we urge further study in this channel with more data particularly because, as discussed above, the mass of the $1^{+-}$ is a useful test of models.

%%%%%%%%%%%%%%%%%%%%%%%%%%%%%%%%%%%%%%%~~~NEW SECTION~~~%%%%%%%%%%%%%%%%%%%%%%%%%%%%%%%%%%%%%%%%%%
%\vspace{0.5em}
{\it \textbf{Conclusions}}.
Recent results from CMS~\cite{CMS:2023owd,CMS:2026tiu} are consistent with the scenario we proposed in our previous work \cite{Anwar:2023fbp}, in which $X(6600)$ and a partner state $X(6400)$ belong to an S-wave multiplet of $cc\c\c$ tetraquarks. The CMS analysis confirms our prediction of $2^{++}$ quantum numbers for $X(6600)$, which also applies in lattice QCD and other theoretical approaches. The CMS data also show evidence for a Breit-Wigner structure around 6400~MeV, with different quantum numbers to other states; we suggest this is the $0^{++}$ partner $X(6400)$. Our predictions for the masses and decays of the scalar and other partner states can be confronted with future experimental data, as a test of theoretical models, in particular to discriminate between quark and effective diquark degrees of freedom.

Our prediction of a scalar state around 6400 MeV stands as the key conclusion of this paper. Establishing an S-wave multiplet of $cc\c\c$ states, either by re-analysing the $\jp \jp$ data or searching $0^{++}$ in the $\eta_c\eta_c$ channel, would be a breakthrough in exotic hadron research and our understanding of exclusively heavy quark exotics.

\begin{acknowledgments}
This work is supported by STFC grant ST/X000648/1 and The Royal Society through Newton International Fellowship.
\end{acknowledgments}

\bibliographystyle{apsrev4-1}

\end{document}